\documentclass[preprint,aps,prc,nofootinbib,showpacs]{revtex4}

\usepackage{epsfig}
\usepackage{graphicx}

\begin{document}

\title{Polarization effects in $e^+ +e^-\rightarrow \bar d+d$ and
determination of time like deuteron form factors}

\author{G. I. Gakh}
\affiliation{ National Science Centre "Kharkov Institute of Physics and Technology",\\ 61108 Akademicheskaya 1, Kharkov,
Ukraine } 

\author{E. Tomasi--Gustafsson}
\affiliation{\it DAPNIA/SPhN, CEA/Saclay, 91191 Gif-sur-Yvette Cedex, 
France}

\author{C. Adamu\v s\v c\'in and S. Dubni\v cka}
\affiliation{Institute of Physics, Slovak Academy of Sciences, Bratislava, Slovakia}

\author{ A.~Z.~Dubni\v ckov\' a}
\affiliation{Department of Theoretical Physics, Comenius University, Bratislava, Slovakia}

\date{\today}

\vspace{0.5cm}
\begin{abstract}
Polarization effects in the reaction $e^++e^-\rightarrow \bar d+d$ have
been investigated for the case of longitudinally polarized electron
beam and arbitrary polarization of the produced deuteron, with the aim of a determination of the time-like complex deuteron electromagnetic form factors. 
General expressions of polarization observables are derived and numerical estimations  have been carried out by means of various models of deuteron electromagnetic form factors,  for kinematical conditions near threshold.
\end{abstract}

\maketitle
 
\section{Introduction}
\hspace{0.7cm}

The electromagnetic form factors (FFs) of hadrons and nuclei provide important
information about the structure and internal dynamics of these systems.
Recent progress in electron--scattering experiments allowed to
measure not only the cross sections but also various polarization observables
in the region of the momentum transfers where these data can help to
discriminate between different theoretical predictions.

The deuteron, the only bound two--nucleon system, is one of the fundamental
systems of nuclear physics. Accordingly, many studies, both experimental and
theoretical, have been devoted to it. Of particular interest today is the
degree to which the deuteron can be understood as a system of two nucleons
interacting via the known nucleon--nucleon interaction.

When addressing, more specifically, to the electromagnetic properties of the
deuteron, the main question concerns the reliability to predict the three
deuteron FFs starting from the calculated deuteron wave function and
nucleon FFs known from electron--nucleon scattering. At low momentum
transfers, predictions and data agree quite well when accounting for one--body
terms only, whereas at the higher momentum transfers, two--body contributions are
known to be important. Whether quark degrees of freedom do need to be taken
explicitly into account, is still a matter of debate. A status of the
experimental and theoretical research of the deuteron can be found in recent 
reviews \cite{GG,S}.

Elastic electron--deuteron scattering has been investigated in
many experiments, and cross section data today covers a large
range of momentum transfers (see review \cite{S}). Some of these
data obviously are not very precise, other data, mainly of more
recent origin, have reached accuracies down to the $1$ \% level.
During the last years, it has become possible to
measure not only cross sections, but also spin observables, due to the
developments of polarized electron beams, polarized deuteron targets and
polarimeters. The
knowledge of these spin observables is unavoidable, if one wants to
separate the contributions of the different multipolarities to
the $A(Q^2)$ structure function. On the side of experiment, good
progress has been made. In particular, recent polarization data for electron--deuteron elastic scattering allowed the individual determination of the deuteron charge and quadrupole FFs up to a value of the momentum transfer squared $Q^2$= 1.8 GeV$^2$.

The deuteron charge FF $G_C$ is particularly interesting for the
understanding the deuteron structure, beyond the impulse approximation. $G_C$ displays a node at $Q^2$=0.7 GeV$^2$, and the position of this node is especially sensitive to the ingredients of the models, in particular meson--exchange currents. 

The experimental investigation of deuteron FFs should help to determine the region where it is necessary to introduce explicitly  quark and gluon degrees of freedom, for a correct description of the deuteron. At present, as it was shown in Ref. \cite{RT}, the overall experimental results on elastic electron--deuteron cross sections are not consistent with pQCD predictions. The best global descriptions of the existing deuteron data are based on impulse approximation (including eventually relativistic corrections, meson exchange currents, $\Delta$ isobars...).

The interaction of electrons with deuterons is usually assumed to occur through the exchange of a virtual photon (one--photon exchange approximation) due to the smallness of the electromagnetic fine structure constant, which suppress two -or more- photon exchange. However, a few decades ago it was suggested \cite{DPh} that the two--photon exchange mechanism may be significant in the region of large momentum transfer. More recently, the possible contribution of two--photon exchange to the elastic electron--deuteron scattering was discussed in Ref. \cite{RT}. 

As for the nucleon, the knowledge of electromagnetic FFs in the time--like (TL) region of momentum transfer can give additional important information about the internal composite structure of the hadron. Measurements are certainly more difficult  in the deuteron case, as shown in Ref. \cite{Du91}, where the total cross section of the reaction $e^-+e^+\rightarrow d + \bar d$ was predicted up to $q^2~=~30$ GeV$^2$, using a model of deuteron FFs based on an extension of the vector--meson--dominance model (VMD) of the electromagnetic hadron interactions. However, other mechanisms, as the presence of a two--photon contribution, could favor a larger cross section.

After the challenging discovery of antideuteron \cite{Ma65}, which established the existence of nuclear antimatter,  the production of antideuteron was recorded in different reactions. Very recently, the production of deuterons and antideuterons in $Au+Au$ collisions
has been reported by the PHENIX experiment at RHIC \cite{Aea} and interpreted in terms of coalescence model. It was found
that the spectra of $d$ and $\bar d$ decrease less steeply than $p$
($\bar p$) spectra. The cross section for $\bar d$ photoproduction was also measured
at HERA at $W_{\gamma p}=200 $ GeV \cite{A1ea}. The production of $\bar d$
in $e^+e^-$--annihilation at $W=10$ GeV was also measured at DORIS II storage
ring \cite{A2ea}.

In the present paper we calculate the polarization observables in the
reaction
\begin{equation}\label{eq:eq1}
e^-(k_1)+e^+(k_2)\rightarrow d(p_1) + \bar d(p_2).
\end{equation}
where the momenta of the particles are indicated in brackets.

We consider the case of unpolarized and longitudinally--polarized electron
beam with production of vector-- and tensor--polarized deuterons. The expressions of polarization observables are given in terms of the deuteron
electromagnetic FFs. Due to final state interaction, FFs are complex
functions of the variable $q^2$. Nevertheless, not all models of FFs, which are mainly parametrizations built for the space-like region, can be consistently applied to the TL region and give origin to an imaginary part. However, numerical estimations for the cross section and polarization observables are tentatively given, on the base of the analytic continuation of existing parameterizations of the deuteron FFs,  similarly to the case of nucleon FFs in Ref. \cite{ETG05}.

\section{Polarization observables}

In the one-photon approximation, the differential cross section of the reaction (\ref{eq:eq1}) in terms of the leptonic
$L_{\mu\nu}$ and hadronic $W_{\mu\nu}$ tensors contraction (in the Born
approximation we can neglect the electron mass) is written as
\begin{equation}\label{eq:eq2}
\frac{d\sigma}{d\Omega } = \frac{\alpha^2\beta }{4q^2}\frac{
L_{\mu\nu}W_{\mu\nu}}{q^4},
\end{equation}
where $\alpha=1/137$ is the electromagnetic constant, $\beta =\sqrt{1-4M^2/q^2}$ is the deuteron velocity in the reaction center of mass system (CMS), M is the deuteron mass and $q$ is the four momentum of the virtual
photon, $q=k_1+k_2=p_1+p_2$ (note that the cross section is not averaged over
the spins of the initial beams).

The leptonic tensor (for the case of longitudinally polarized electron
beam) is
\begin{equation}\label{eq:eq3}
L_{\mu\nu}=-q^2g_{\mu\nu}+2(k_{1\mu}k_{2\nu}+k_{2\mu}k_{1\nu}) +
2i\lambda \varepsilon_{\mu\nu\sigma\rho}k_{1\sigma}k_{2\rho}\ , 
\end{equation}
where $\lambda$ is the degree of the beam polarization (further we assume that
the electron beam is completely polarized and consequently $\lambda=1$). 

The hadronic tensor can be expressed via the nucleon electromagnetic current
$J_{\mu}$, describing the transition $\gamma ^*\rightarrow \bar dd$, as
\begin{equation}\label{eq:eq4}
W_{\mu\nu} = J_{\mu}J^*_{\nu}\ .
\end{equation}
As the deuteron is a spin--one nucleus, its electromagnetic current is
completely described by three FFs. Assuming the P-- and C--invariance
of the hadron electromagnetic interaction this current can be written
as \cite{AR} 
\begin{equation}\label{eq:eq5}
J_{\mu}=(p_1-p_2)_{\mu}[-G_1(q^2)U_1^*\cdot U_2^*+\frac{G_3(q^2)}{M^2}
(U_1^*\cdot q U_2^*\cdot q-
\frac{q^2}{2}U_1^*\cdot U_2^*)]-G_2(q^2)(U_{1\mu}^*U_2^*\cdot q-
U_{2\mu}^*U_1^*\cdot q),
\end{equation}
where $U_{1\mu}~(U_{2\mu})$ is the polarization four-vector describing the spin one deuteron (antideuteron), and $G_i(q^2)~ (i=1, 2, 3)$ are the deuteron electromagnetic FFs. The FFs $G_i(q^2)$ are complex functions of the variable
$q^2$ in the region of the TL momentum transfer ($q^2>0$).  They are related to the standard deuteron electromagnetic FFs:  $G_C$ 
(charge monopole), $G_M$ (magnetic dipole) and $G_Q$ (charge quadrupole) by
\begin{equation}\label{eq:eq6}
G_M=-G_2, \ G_Q=G_1+G_2+2G_3, \ \
G_C=-\frac{2}{3}\tau (G_2-G_3)+ (1-\frac{2}{3}\tau )G_1, \
\ \tau=\frac{q^2}{4M^2}\ .
\end{equation}
The standard FFs have the following normalizations:
\begin{equation}\label{eq:eq7}
G_C(0)=1\ , \ \  G_M(0)=(M/m_n)\mu_d\ , \ \ G_Q(0)=M^2Q_d\ ,
\end{equation}
where $m_n$ is the nucleon mass, $\mu_d=0.857(Q_d=0.2859$ fm$^2$) is deuteron magnetic (quadrupole) moment.

When calculating the expression for the hadron tensor $W_{\mu\nu}$ in terms
of the deuteron electromagnetic FFs, using the explicit form of the
electromagnetic current (\ref{eq:eq5}), the spin--density matrices of the deuteron and antideuteron are 
\begin{equation}\label{eq:eq8}
U_{1\mu}U^*_{1\nu}=-\left(g_{\mu\nu}-\frac{p_{1\mu}
p_{1\nu}}{M^2}\right) +\frac{3i}{2M}
 \varepsilon_{\mu\nu\rho\sigma}s_{\rho} p_{1\sigma}
+3Q_{\mu\nu} \ , \ \ 
U_{2\mu}U^*_{2\nu}=-\left(g_{\mu\nu}-\frac{p_{2\mu}
p_{2\nu}}{M^2}\right)\ ,
\end{equation}
if the deuteron polarization is measured and the antideuteron polarization is
not measured. Here $s_{\mu}$ and $Q_{\mu\nu}$ are the deuteron
polarization four vector and quadrupole tensor, respectively. The four vector of the deuteron vector polarization $s_{\mu}$ and the deuteron quadrupole--polarization tensor $Q_{\mu\nu }$ satisfy the following conditions:
 $$s^2=-1,~ sp_1=0,~Q_{\mu\nu}=Q_{\nu\mu}, \ \ Q_{\mu\mu}=0, ~
p_{1\mu}Q_{\mu\nu}=0\ .$$

Taking into account Eqs. (\ref{eq:eq4}), (\ref{eq:eq5}) and (\ref{eq:eq8}), the hadronic tensor in the 
general case can be written as the sum of three terms
\begin{equation}\label{eq:eq9}
W_{\mu\nu}=W_{\mu\nu}(0)+W_{\mu\nu}(V)+W_{\mu\nu}(T),
\end{equation}
where $W_{\mu\nu}(0)$ corresponds to the case of unpolarized deuteron and
$W_{\mu\nu}(V) (W_{\mu\nu}(T))$ corresponds to the case of the
vector (tensor) polarized deuteron. The explicit form of these terms is:

\noindent\underline{ - the unpolarized term $W_{\mu\nu}(0)$:}

$$W_{\mu\nu}(0)=W_1(q^2)\tilde{g}_{\mu\nu}+\frac{W_2(q^2)}{M^2}
\tilde{p}_{1\mu}\tilde{p}_{1\nu} \ , \
\tilde{g}_{\mu\nu}=g_{\mu\nu}-\frac{q_{\mu}q_{\nu}}{q^2}\ , \ \
\tilde{p}_{1\mu}=p_{1\mu}-\frac{p_1q}{q^2}q_{\mu} \ , $$
\begin{equation}\label{eq:eq10, structure functions}
W_1(q^2)=8M^2\tau (1-\tau )|G_M|^2, \ \ 
W_2(q^2)=12M^2(|G_C|^2-\frac{2}{3}\tau |G_M|^2+\frac{8}{9}\tau ^2|G_Q|^2).
\end{equation}

\noindent\underline{ - the term for vector polarization $W_{\mu\nu}(V)$:}

\begin{equation}\label{eq:eq11, structure functions}
W_{\mu\nu}(V)=\frac{i}{M}S_1(q^2)\varepsilon_{\mu\nu\sigma\rho} s_{\sigma}q_{\rho}+
\frac{i}{M^3}S_2(q^2)[\tilde{p}_{1\mu}
\varepsilon_{\nu\alpha\sigma\rho} s_{\alpha}q_{\sigma}p_{1\rho}-\tilde{p}_{1\nu}
\varepsilon_{\mu\alpha\sigma\rho} s_{\alpha}q_{\sigma}p_{1\rho}]+
\end{equation}
$$+\frac{1}{M^3}S_3(q^2)[\tilde{p}_{1\mu}
\varepsilon_{\nu\alpha\sigma\rho} s_{\alpha}q_{\sigma}p_{1\rho}+\tilde{p}_{1\nu}\varepsilon_{\mu\alpha\sigma\rho} s_{\alpha}q_{\sigma}p_{1\rho}], \ $$
$$S_1(q^2)=-3M^2(\tau -1)|G_M|^2, \
S_2(q^2)=3M^2[|G_M|^2-2Re(G_C-\frac{\tau }{3}G_Q)G_M^*], \ $$
$$S_3(q^2)=6M^2Im(G_C-\frac{\tau }{3}G_Q)G_M^*. \ $$

\noindent\underline{ - the term for tensor polarization $W_{\mu\nu}(T)$:}

\begin{equation}\label{eq:eq12}
W_{\mu\nu}(T)=V_1(q^2)\bar Q\tilde{g}_{\mu\nu}+V_2(q^2)\frac{\bar Q}{M^2}
\tilde{p}_{1\mu}\tilde{p}_{1\nu}+
\end{equation}
$$+V_3(q^2)(\tilde{p}_{1\mu}\widetilde{Q}_{
\nu}+\tilde{p}_{1\nu}\widetilde{Q}_{\mu})+
V_4(q^2)\widetilde{Q}_{\mu\nu}\ +
iV_5(q^2)(\tilde{p}_{1\mu}\widetilde{Q}_{
\nu}-\tilde{p}_{1\nu}\widetilde{Q}_{\mu}),  $$
where
$$\widetilde{Q}_{\mu}=Q_{\mu\nu}q_{\nu}-\frac{q_{\mu}}{q^2}\bar
{Q} \ , \ \ \widetilde{Q}_{\mu}q_{\mu}=0\ , $$
\begin{equation}\label{eq:eq13}
\widetilde{Q}_{\mu\nu}=
Q_{\mu\nu}+\frac{q_{\mu}q_{\nu}}{q^4}\bar Q-
\frac{q_{\nu}q_{\alpha}}{q^2}Q_{\mu\alpha}-
\frac{q_{\mu}q_{\alpha}}{q^2}Q_{\nu\alpha}\ , \
\widetilde{Q}_{\mu\nu}q_{\nu} = 0, \ \bar Q=Q_{\alpha\beta}q_{\alpha}q_{\beta}.
\end{equation}

The tensor structure functions $V_i(q^2)$ are combinations of deuteron FFs as follows:
\begin{equation}\label{eq:eq14}
V_1(q^2)=-3|G_M|^2, \
V_2(q^2)=3\left [|G_M|^2+\frac{4}{1-\tau }Re(G_C-\frac{\tau }{3}G_Q
-\tau G_M)G_Q^*\right ], \
\end{equation}
$$V_3(q^2)=-6\tau \left [|G_M|^2+2ReG_QG_M^* \right ], \
V_4(q^2)=-12M^2\tau (1-\tau )|G_M|^2, \
V_5(q^2)=-12\tau Im (G_QG_M^*). \ $$

Using the definitions of the cross--section (\ref{eq:eq2}), leptonic (\ref{eq:eq3}) and
hadronic (\ref{eq:eq9}) tensors, one can easily derive the expression for the unpolarized
differential cross section in terms of the structure functions $W_{1,2}$
(after averaging over the spins of the initial particles)
\begin{equation}\label{eq:eq15}
\frac{d\sigma^{un}}{d\Omega }=\frac{\alpha^2\beta }{4q^4} \left \{
-W_1(q^2)+\frac{1}{2}W_2(q^2) \left [\tau -1-\frac{(u-t)^2}{4M^2q^2}\right ]\right \}\ ,
\end{equation}
where $t=(k_1-p_1)^2$, $ u=(k_1-p_2)^2.$

In the reaction CMS this expression can be written as
\begin{equation}\label{eq:eq16}
\frac{d\sigma^{un}}{d\Omega }=\frac{\alpha^2\beta ^3}{4q^2}D, \
D=\tau (1+\cos^2\theta )|G_M|^2+\frac{3}{2}\sin^2\theta \left (|G_C|^2+
\frac{8}{9}\tau ^2|G_Q|^2 \right ),
\end{equation}
where $\theta $ is the angle between the momenta of the deuteron (${\vec p}$)
and the electron beam (${\vec k}$). 
Integrating the expression (16) with respect to the deuteron angular
variables one obtains the following formula for the total cross section of
the reaction (\ref{eq:eq1})
\begin{equation}\label{eq:i1}
\sigma_{tot}(e^+e^-\to \bar dd)=
\frac{\pi\alpha ^2\beta^3}{3q^2}\biggl [3|G_{C}|^2+4\tau
(|G_{M}|^2+\frac{2}{3}\tau |G_{Q}|^2)\biggr ]. \
\end{equation}

One can define also an angular asymmetry, $R$, with respect to the
differential cross section measured at $\theta =\pi /2, \sigma_0$
\begin{equation}\label{eq:i2}
\frac{d\sigma^{un}}{d\Omega }=\sigma_0(1+Rcos^2\theta ),
\end{equation}
where $R$ can be expressed as a function of the deuteron FFs
\begin{equation}\label{eq:i3}
R=\frac{2\tau (|G_{M}|^2-\frac{4}{3}\tau |G_{Q}|^2)-3|G_{C}|^2}
{2\tau (|G_{M}|^2+\frac{4}{3}\tau |G_{Q}|^2)+3|G_{C}|^2}.
\end{equation}
This observable should be sensitive to the different underlying
assumptions on deuteron FFs; therefore, a precise measurement of
this quantity, which does not require polarized particles, would be
very interesting.

One can see that, as in the space-like (SL) region,  the measurement of the
angular distribution of the outgoing deuteron determines the modulus of the
magnetic form factor, but the separation of the charge and quadrupole form
factors requires the measurement of polarization observables \cite{ACG}. The
outgoing--deuteron polarization can be measured in a secondary analyzing
scattering \cite{ACG}. For vector polarization up to a few GeV, an inclusive measurement on a carbone target as $d+C\to one$ $charged$ $particle$ $+X$ is sufficient, when the charged protons from deuteron break up are eliminated with help of an absorber \cite{Bo90}. For tensor polarization, however, only exclusive reactions as elastic $d+p$ scattering \cite{HYPOM} or charge exchange \cite{Polder} give sufficient efficiency and analyzing powers.

As it was shown in Ref. \cite{DDR}, a nonzero phase difference between FFs of two baryons (with 1/2 spins) leads to the T--odd single--spin
asymmetry normal to the scattering plane in the baryon--antibaryon production
$e^+e^-\rightarrow B\bar B$. It is 
more convenient to derive polarization observables in CMS. When considering the
polarization of the final particle, we choose a reference system  with the  $z$ axis along the
momentum of this particle (in our case it is ${\vec p}$). The $y$ axis is
normal to the reaction plane in the direction of ${\vec k}\times {\vec p}$; $x$, $y$ and $z$ form a right--handed coordinate system.

The cross section can be written, in the general case, as the sum of unpolarized and polarized terms, corresponding to the different polarization states and polarization directions of the incident and scattered particles:
\begin{equation}\label{eq:eq21}
\displaystyle\frac{d\sigma}{d\Omega}=
\displaystyle\frac{d\sigma^{un}}{d\Omega}
\left [1+P_y+\lambda P_x+\lambda P_z+ P_{zz}R_{zz}+
P_{xz}R_{xz}+P_{xx}(R_{xx}-R_{yy}) +\lambda P_{yz}R_{yz}\right ],
\end{equation}
where $P_i$ ($P_{ij}$), $i,j=x,y,z$ are the components of the polarization vector (tensor) of the outgoing deuteron, $R_{ij}$, $i,j=x,y,z$ the components of the quadrupole polarization tensor of the outgoing deuteron
$Q_{\mu\nu}$, in its rest system and $\displaystyle\frac{d\sigma^{un}}{d\Omega}$ is the differential cross section for the unpolarized case.

The degree of longitudinal polarization of the electron beam, $\lambda$, is explicitly indicated, in order to stress the origin of the specific polarization observables.

Let us consider the different polarization observables and give their expression in terms of the deuteron FFs.

\begin{itemize}
\item 

The vector polarization of the outgoing deuteron, $P_y$, which does not require polarization in the initial state is
\begin{equation}\label{eq:eq17}
P_y=-\frac{3}{2}\sqrt{\tau }\sin(2\theta )Im
\left [\left (G_C-\frac{\tau }{3}G_Q \right ) G_M^*\right ]/ D.
\end{equation}

\item
The part of the differential cross section that depends on the tensor
polarization can be written as follows
\begin{eqnarray}
&\displaystyle\frac{d\sigma_T}{d\Omega}&=
\frac{d\sigma_{zz}}{d\Omega}R_{zz}+\frac{d\sigma_{xz}}{d\Omega}R_{xz}+
\frac{d\sigma_{xx}}{d\Omega}(R_{xx}-R_{yy}),
\\
&\displaystyle\frac{d\sigma_{zz}}{d\Omega}&=\frac{\alpha ^2\beta ^3}{4q^2}
\frac{3\tau }{4}\left [(1+\cos^2\theta )|G_M|^2+8\sin^2\theta 
\left (
\frac{\tau }{3}|G_Q|^2-Re(G_CG_Q^*)\right ) \right], \\
&\displaystyle\frac{d\sigma_{xz}}{d\Omega}&=-\frac{\alpha ^2\beta ^3}{4q^2}
3\tau ^{3/2}\sin(2\theta )Re(G_QG_M^*),
\\
&\displaystyle\frac{d\sigma_{xx}}{d\Omega}&=-\frac{\alpha ^2\beta ^3}{4q^2}
\frac{3\tau }{4}\sin^2\theta |G_M|^2, 
\end{eqnarray}

\item 
Let us consider now the case of a longitudinally polarized electron beam.
The other two components of the deuteron vector polarization ($P_x$, $P_z$)
require the initial particle polarization and are
\begin{equation}\label{eq:eq20}
P_x=-3\frac{\sqrt{\tau }}{D}\sin\theta Re \left (G_C-\frac{\tau }{3}G_Q \right )G_M^*, \
P_z=\frac{3\tau }{2D}\cos\theta |G_M|^2.
\end{equation}
\end{itemize}
From angular momentum and helicity conservations it follows that the sign of the deuteron polarization component $P_z$ in the forward direction 
($\theta =~0$) must coincide with the sign of the electron beam polarization. This requirement is satisfied by Eq. (\ref{eq:eq20}).

A possible nonzero phase difference between the deuteron FFs leads
to another T--odd polarization observable proportional to the $R_{yz}$
component of the tensor polarization of the deuteron.
The part of the differential cross section that depends on the correlation
between the longitudinal polarization of the electron beam and the deuteron
tensor polarization can be written as follows
\begin{equation}\label{eq:eq21a}
\frac{d\sigma_{\lambda T}}{d\Omega}=\frac{\alpha ^2\beta ^3}{4q^2}
6\tau ^{3/2}\sin\theta Im(G_MG_Q^*)R_{yz}.
\end{equation}

The deuteron FFs in the TL region are complex functions. In 
the case of unpolarized initial and final particles, the differential cross
section depends only on the squared modulus $|G_M|^2$ and on the combination 
$G = |G_C|^2+\frac{8}{9}\tau ^2|G_Q|^2.$ So, the measurement of the angular 
distribution allows one to determine $|G_M|$ and the quantity 
$G$, as in the elastic electron--deuteron scattering. 

Let us discuss which information can be obtained by measuring the polarization observables derived above. Three relative phases exist for three FFs, which we note  
as follows: $\alpha_1=\alpha_M-\alpha_Q,$ $\alpha_2=\alpha_M-\alpha_C,$ and 
$\alpha_3=\alpha_Q-\alpha_C,$ where $\alpha_M=ArgG_M,$ $\alpha_C=ArgG_C,$ and 
$\alpha_Q=ArgG_Q.$ These phases are important characteristics of FFs in the TL region since they result from the strong interaction between final particles.

Let us consider the ratio of the polarizations $P_{yz}$ (let us remind that it requires a 
longitudinally polarized electron beam) and $P_{xz}$ (when the electron beam is 
unpolarized). One finds:
\begin{equation}\label{eq:eq22}
R_1=\frac{P_{xz}}{P_{yz}}=-\cos\theta \cot\alpha_1.
\end{equation}
So, the measurement of this ratio gives us information about the relative phase $\alpha_1$. The measurement of another ratio of polarizations, $R_2=P_{xz}/P_{xx}$ gives us information about the quantity $|G_Q|$:
\begin{equation}\label{eq:eq23}
R_2=\frac{P_{xz}}{P_{xx}}=8\sqrt{\tau }\cot\theta \cos\alpha_1\frac{|G_Q|}{|G_M|}.
\end{equation}
This allows one to obtain the modulus of the charge FF,  $|G_C|$, from the quantity $G$, known from the measurement of the differential cross section. The measurement of a third ratio
\begin{equation}\label{eq:eq24}
R_3=\frac{P_{y}}{P_{x}}=-\cos\theta \frac{\sin\alpha_2-r\sin\alpha_1}
{\cos\alpha_2-r\cos\alpha_1}, \ \ r=\frac{\tau }{3}\frac{|G_Q|}{|G_C|}
\end{equation}
allows to determine the phase difference $\alpha_2$. And at last, if we measure 
the ratio of the polarizations $P_{zz}$ and $P_{xx}$
\begin{equation}\label{eq:eq25}
R_4=\frac{P_{zz}}{P_{xx}}=-\frac{1}{\sin^2\theta}
\left [1+\cos^2\theta +8\sin^2\theta 
\frac{|G_C||G_Q|}{|G_M|^2}(r-\cos\alpha_3) \right ]
\end{equation}
we can obtain information about the third phase difference $\alpha_3$. Moreover,  one can verify the relation:
$$\alpha_3=\alpha_2-\alpha_1.$$

Thus, the measurement of these polarization observables allows to fully determine the deuteron FFs in TL region. 

Note that using the ratio of two  polarization components that are simultaneously measured, greatly reduces systematic uncertainties. It is not 
necessary to know neither the beam polarization or the polarimeter analyzing 
power, since both of these quantities cancel in the ratio. 

This procedure can be considered as the generalization of the polarization method proposed almost four decades ago \cite{Re68}, which could be applied only recently to elastic electron proton scattering \cite{JG00}.

Let us note here that, in principle, one should take into account the problem of the two--photon--exchange contribution, which, as mentioned in the Introduction, may become important at large momentum transfer, as it is expected that the reactions mechanisms are similar for the crossed channel (\ref{eq:eq1}). As it was shown in Ref. \cite{Pu61}, if the detection of  the 
final particles does not distinguish between deuteron and antideuteron, then the interference between one--photon and two--photon amplitudes does not contribute to the cross section of the reaction (\ref{eq:eq1}).

\section{Numerical estimations}

In the previous section, the expressions of cross section and polarization observables have been given, in terms of the deuteron FFs. Numerical estimations require the knowledge of such FFs, in TL region. Due to the hermiticity of the electromagnetic current, FFs are real in the SL region, and complex in the TL region. At our knowledge, most of the existing parametrizations of these FFs are phenomenological fits to SL data, and are useful for different estimations and to plan corresponding experiments in that kinematical region. However, their analytical expressions were not built to obey fundamental properties of FFs. For example, their extension to the TL region does not induce any phase (i.e., the imaginary part of FFs is equal to zero).

Recent work in this direction \cite{www} describes three different parametrizations of deuteron FFs describing the world data. The first one (Parametrization I) is a sum of inverse polynomial terms, where the first node of the corresponding FFs is introduced as a global multiplicative term. The number of free parameters,  necessary to obtain $\chi^2/ndf=1.5 $, was 18. 

The second parametrization is based on a previous work 
\cite{Ko95}. It is an attempt to find a global description based on the vector dominance model,  satisfying the asymptotic conditions predicted by QCD at large momentum transfer, and leads to a 12 parameters fit.

The third parametrization is a sum of gaussians, with some physical constraints on the parameters, which are the width and the position of the maximum of the gaussians. In total the parametrization contains 33 parameters for $\chi^2/ndf=1.5 $.

In Ref. \cite{ETG06} a generalization of the nucleon model from Ref. \cite{Ia73} has been successfully applied to the deuteron case. Besides the fact that the VMD model \cite{Ia73} satisfies by construction some of the basic properties of FFs, its extension to the TL region is straightforward \cite{Wa04}.

The basic idea of this parametrization is the presence of two components in the hadron structure: an intrinsic structure, very compact, characterized by a dipole (monopole) $q^2$ dependence and a meson cloud, which contains only the $\rho$, $\phi$ and $\omega$ (not the $\rho$) contributions, in the nucleon (deuteron) case. A very good description of all known data on deuteron electromagnetic FFs has been obtained, with as few as six free parameters and few evident physical constraints.  

In principle, all these parametrizations are not predictive outside the kinematical domain where the experimental data have been fitted. Therefore, the extrapolation to TL region is just given for illustrative purposes. We give the predictions from one of the parametrizations from Ref. \cite{www} (Parametrization I), and of the model from \cite{ETG06}. Note that the analytical form of all three parametrizations in \cite{www} is such that only real terms are present in TL region. An imaginary part arises naturally from the analytical continuation of model \cite{ETG06}, (for $q^2\to -q^2)$ due to the not integer  nature of the exponent of the intrinsic part. Finite widths for the $\phi$ and $\omega$ meson contributions would also give rise to complexity, but it was not necessary to introduce them, for obtaining a good description of data in SL region.

We also consider an updated version of the model \cite{Du91}, based on unitarity and analyticity \cite{Ad06}.

\begin{figure}
\mbox{\epsfxsize=14.cm\leavevmode \epsffile{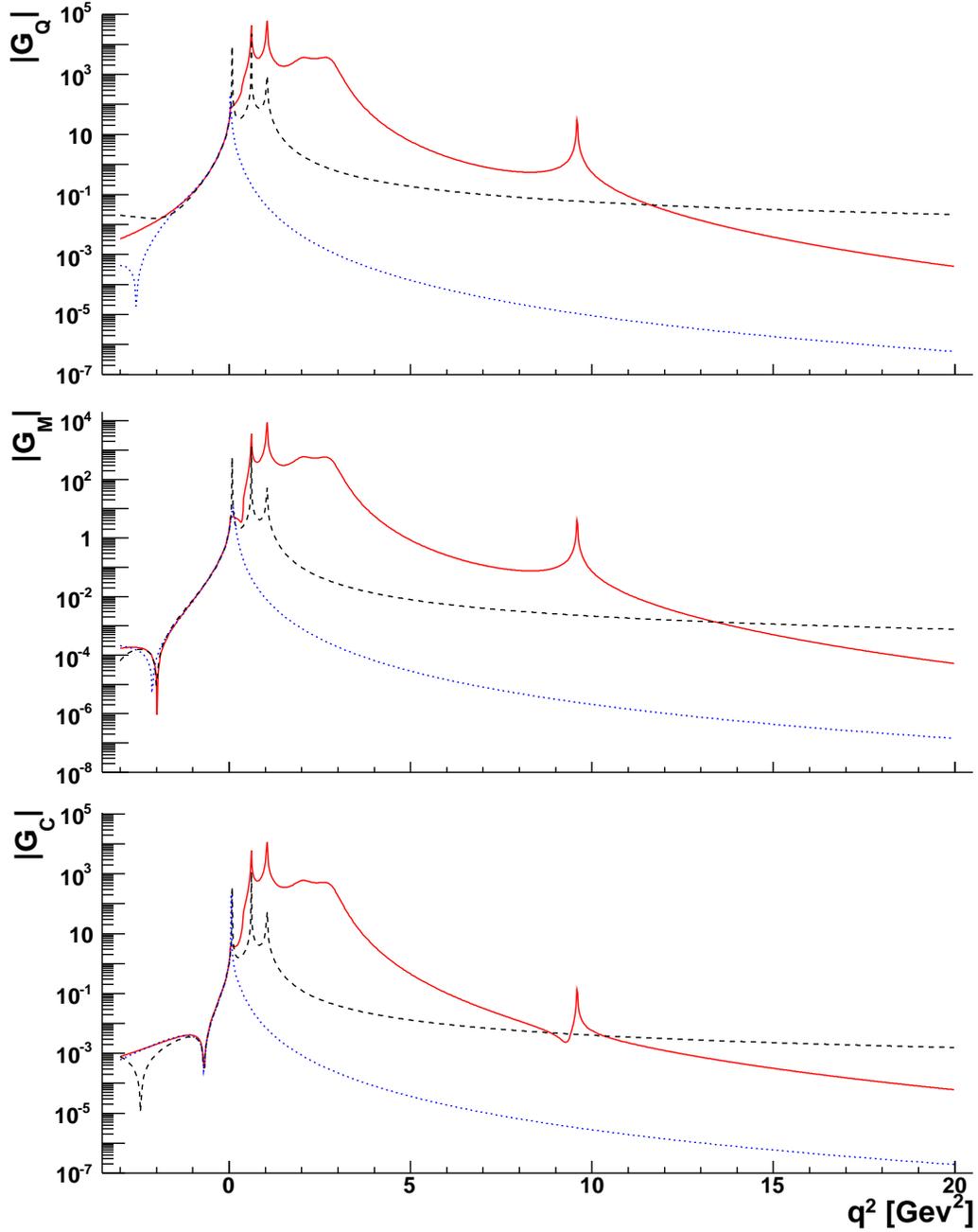}}
\vspace*{.2 truecm}
\caption{$q^2$-dependence of the $G_Q$, $G_M$, $G_C$ from top to bottom (moduli): from Ref. \protect\cite{Ad06} (solid line), from Ref. \protect\cite{ETG06} (dashed line), and from Parametrization I from Ref. \protect\cite{www} (dotted line).}
\label{fig:mod}
\end{figure}

\begin{figure}
\mbox{\epsfxsize=14.cm\leavevmode \epsffile{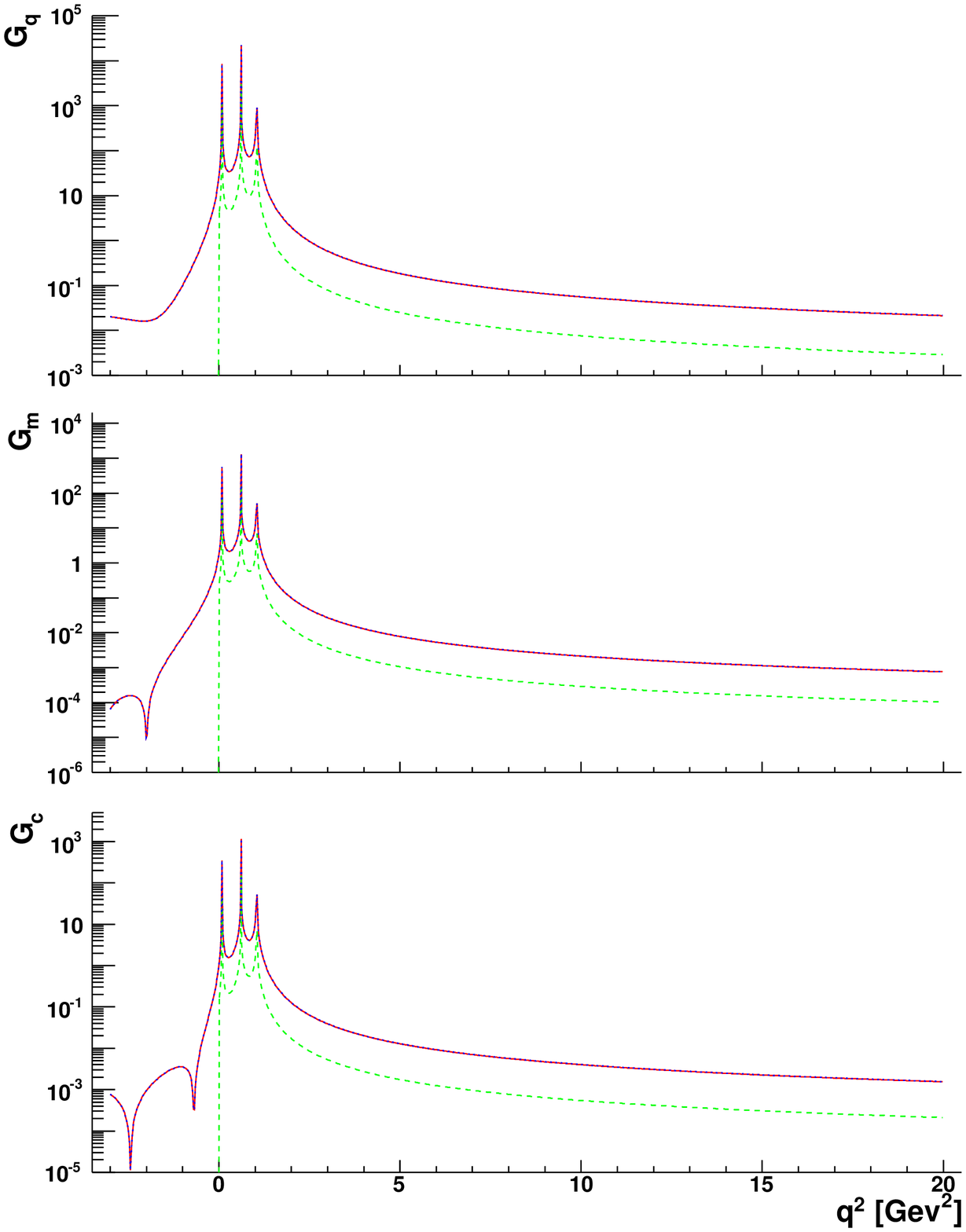}}
\vspace*{.2 truecm}
\caption{$q^2$-dependence of the $G_Q$, $G_M$, $G_C$ from top to bottom: from Ref. \protect\cite{ETG06}: real part (solid line), imaginary part (dashed line).}
\label{fig:ffs}
\end{figure}

The $q^2$ dependence of these models is illustrated in Fig. \ref{fig:mod} for the moduli and in Fig. \ref{fig:ffs}  for the real an imaginary parts of the model from Ref. \cite{ETG06}. One can see that the three models coincide in the SL region, where they are constrained by the experimental data, but, outside this kinematical region, they show very different behavior. In particular,  Parametrization I differs by few order of magnitude. Parametrization I does not show any singularity in TL region. Two poles coincide in TL region, for the models  \cite{Ad06} and \cite{ETG06},  as they correspond to the $\omega$ and $\phi$ contributions. More resonances are built, by construction, in  the model \cite{Ad06}, and occur in the unphysical region. These two models show a similar trend, near the threshold, for the moduli of FFs, however the sign, which is reflected in the relevant polarization observables, may differ.

From Fig. \ref{fig:ffs} one can see that in TL region, FFs from Ref. \cite{ETG06} display an imaginary part which is an order of magnitude smaller than the real part, as a consequence of the exponent of the term corresponding to the intrinsic part. As the model \cite{Ad06} fulfills by construction the unitary condition, its imaginary part starts at the deuteron anomalous threshold, $q^2=1.73 m_{\pi}^2$ $\simeq ~0.02$ GeV$^2$. Concerning the model \cite{ETG06}, the imaginary part is different from zero for $q^2>0.08$ GeV$^2$.

\begin{figure}
\mbox{\epsfxsize=14.cm\leavevmode \epsffile{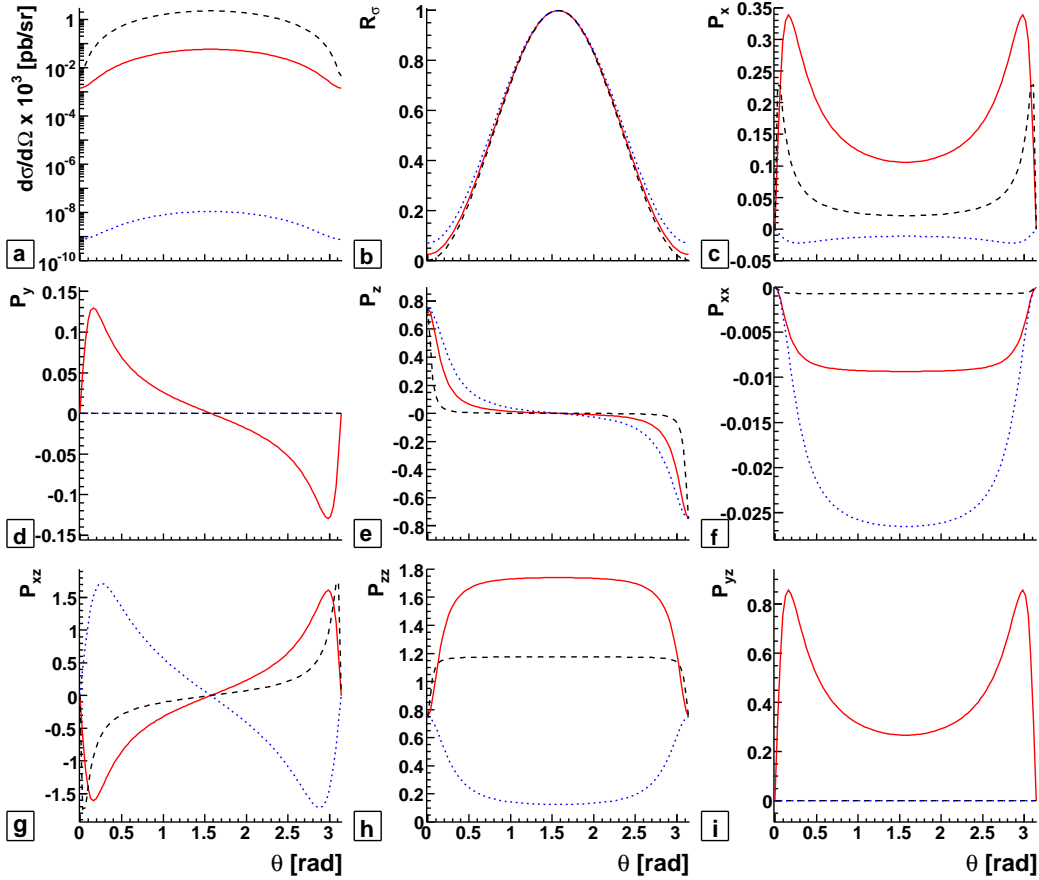}}
\vspace*{.2 truecm}
\caption{Predictions of the different observables, for the considered  parametrizations of deuteron FFs, extrapolated to the TL region. Notations as in Fig. 1.}
\label{Fig:fig1}
\end{figure}

The predictions for the different observables are shown in Fig. \ref{Fig:fig1}, for $E$=1.9 GeV, not far above threshold.  The three parametrizations, as expected, give very different results, especially concerning the predictions for the cross section (Fig. \ref{Fig:fig1}a), which just reflects the differences in the moduli of FFs. In spite of this, the angular distributions are very similar, as it appears from Fig. \ref{Fig:fig1}a, as it is driven by the underlying assumption of the one-photon exchange mechanism.

Evidently, the observables such as $P_y$ (Fig. \ref{Fig:fig1}d) and $P_{yz}$ (Fig. \ref{Fig:fig1}i) vanish, for  parametrization I, as they depend only on the imaginary part, see Eqs. (\ref{eq:eq17}) and (\ref{eq:eq21a}), respectively. 

In the physical region, the angular asymmetry, Eq. (\ref{eq:i3}), is very large in absolute value (over 90\%) and negative, for all the considered models, due to the fact that one FF, $G_Q$, is dominant.

It should be noted that the CMS threshold energy of the reaction $e^++e^-\rightarrow \bar d+d$ is quite large, $E_T= 2M\simeq 3.75$ GeV, which corresponds to $q^2 \simeq 14$ GeV$^2$. There are no data in this momentum range in SL region, which could better constrain models and parametrizations. 

\section{Conclusions}

Polarization observables have been derived for the production of a deuteron antideuteron pair in electron-positron annihilation. Although the cross section of this process is expected to be very small, the search for the corresponding events it is not excluded in future at high luminosity $e^+e^-$ rings.

In TL region, the electromagnetic structure of the deuteron is characterized by three complex FFs. Generalizing the polarization method, successfully applied to $ep$ elastic scattering, we derive the expressions for the relevant observables in terms of the deuteron FFs and indicate the measurements which are necessary for the full determination of the deuteron structure.

Quantitative estimations require the knowledge of the deuteron FFs, in the corresponding kinematical region. Data are absent in the whole TL region, and  also in SL region, at large momentum transfer squared. Therefore, we used the analytical continuations from the SL region of few existing parametrizations and models, keeping in mind that they are poorly constrained in the corresponding SL kinematical region. The results show that polarization effects either vanish or are large and measurable.

The formalism developed here is model independent and based on symmetry properties of electromagnetic and strong interactions. It allows to establish properties of observables that should be satisfied by any model calculation. Moreover, it applies as well to the annihilation reactions involving the production of spin one particles in the final state, such as $e^++e^-\to \rho^++\rho^-$,  $e^++e^-\to \omega^++\omega^-$. The study of these reactions will be the object of a future work.

\section{Acknowledgments}
The authors acknowledge Prof. M. P. Rekalo for enlightening discussions and ideas, without which this paper would not have been realized in the present form.

The Slovak Grant Agency for Sciences VEGA is acknowledged by C.A., S.D., and A. Z. D. for support under Grant N. 2/4099/26.

\section{Appendix}
\hspace{0.7cm}
\setcounter{equation}{0}
\def\theequation{C.\arabic{equation}}

In this Appendix we give useful formulae describing the polarization state
of the deuteron for different cases. For the case of arbitrary
polarization,the deuteron is described by the spin--density
matrix (defined, in the general case, by 8 parameters) which, in the
coordinate representation, has the form
\begin{equation}\label{eq:eqC1}
\rho_{\mu\nu}=-\frac{1}{3}\bigl(g_{\mu\nu}-\frac{p_{\mu}p_{\nu}}{M^2}\bigr)
+\frac{i}{2M}\varepsilon_{\mu\nu\lambda\rho}s_{\lambda}p_{\rho}+ Q_{\mu\nu},
\ \ Q_{\mu\nu}=Q_{\nu\mu}, \ \ Q_{\mu\mu}=0\ , \ \ p_{\mu}Q_{\mu\nu}=0\ ,
\end{equation}
where $p_{\mu }$ is the deuteron four momentum, $s_{\mu}$ and $Q_{\mu\nu}$ are
the deuteron polarization four vector and the deuteron quadrupole polarization
tensor.

In the deuteron rest frame the above formula is written as
\begin{equation}\label{eq:eqC2}
\rho_{ij}=\frac{1}{3}\delta_{ij}-\frac{i}{2}\varepsilon
_{ijk}s_k+Q_{ij}, \ ij=x,y,z.
\end{equation}
This spin--density matrix can be written in the helicity representation
using the following relation
\begin{equation}\label{eq:eqC3}
\rho_{\lambda\lambda'}=\rho_{ij}e_i^{(\lambda )*}e_j^{(\lambda')}, \
\lambda ,\lambda'=+,-,0,
\end{equation}
where $e_i^{(\lambda )}$ are the deuteron spin functions which have the
deuteron spin projection $\lambda $ on to the quantization axis (z axis).
They are
\begin{equation}\label{eq:eqC4}
e^{(\pm )}=\mp \frac{1}{\sqrt{2}}(1,\pm i,0), \
e^{(0)}=(0,0,1).
\end{equation}
The elements of the spin--density matrix in the helicity representation
are related to the ones in the coordinate representation by such a way
\begin{equation}\label{eq:eqC5}
\rho _{\pm\pm}=\frac{1}{3}\pm \frac{1}{2}s_z-\frac{1}{2}Q_{zz}, \ \ 
\rho_{00}=\frac{1}{3}+Q_{zz}, \ \ 
\rho_{+-}=-\frac{1}{2}(Q_{xx}-Q_{yy})+iQ_{xy}, \
\end{equation}
$$\rho_{+0}=\frac{1}{2\sqrt{2}}(s_x-is_y)-
\frac{1}{\sqrt{2}}(Q_{xz}-iQ_{yz}), \ \ 
\rho_{-0}=\frac{1}{2\sqrt{2}}(s_x+is_y)+
\frac{1}{\sqrt{2}}(Q_{xz}+iQ_{yz}), \ $$
$$\rho_{\lambda\lambda'}=
(\rho_{\lambda'\lambda})^* .  \  $$

To obtain these relations we use $Q_{xx}+Q_{yy}+Q_{zz}=0.$

When the deuteron is used as a target, the spin matrix is diagonal, and the polarization state is described by the population numbers
$n_+, \ $ $n_- $ and $n_0 $. Here
$n_+, \ $ $n_- $ and $n_0 $ are the fractions of the atoms with
the nuclear spin projection on to the quantization axis
$m=+1, \ $ $m=-1$ and $m=0,$
respectively. If the spin--density matrix is normalized to 1, i.e.
$Tr \rho =1$, then we have $n_++n_-+n_0=1$. Thus, the polarization state of
the deuteron target is defined in this case by two parameters, called 
V (vector) and T (tensor) polarizations
\begin{equation}\label{eq:eqC6}
V=n_+-n_-, \ \ T=1-3n_0.
\end{equation}
Using the definitions for the quantities $n_{\pm ,0}$
\begin{equation}\label{eq:eqC7}
n_{\pm }=\rho_{ij}e_i^{(\pm )*}e_j^{(\pm )}, \ \ 
n_0=\rho_{ij}e_i^{(0)*}e_j^{(0)},
\end{equation}
we have the following relation between $V$ and $T$ parameters and parameters
of the spin--density matrix in the coordinate representation 
(with the quantization axis directed along the $z$ axis)
\begin{equation}\label{eq:eqC8}
n_0=\frac{1}{3}+Q_{zz}, \ \ n_{\pm }=\frac{1}{3}\pm \frac{1}{2}s_z-
\frac{1}{2}Q_{zz},
\end{equation}
or
\begin{equation}\label{eq:eqC9}
T=-3Q_{zz}, \ \ V=s_z.
\end{equation}

\end{document}